\documentclass[10pt,longbibliography, amsmath, twocolumn, amssymb, aps, prl]{revtex4-2}
\usepackage{xr}
\usepackage{amsmath}
\makeatletter
\newcommand*{\addFileDependency}[1]{
  \typeout{(#1)}
  \@addtofilelist{#1}
  \IfFileExists{#1}{}{\typeout{No file #1.}}
}
\makeatother
\usepackage{lineno}

\newcommand*{\myexternaldocument}[1]{%
    \externaldocument{#1}%
    \addFileDependency{#1.tex}%
    \addFileDependency{#1.aux}%
}
\myexternaldocument{Supplemental}

\usepackage{graphicx}   
\usepackage{dcolumn}  
\usepackage{bm}          
\usepackage{xcolor}    

\usepackage{hyperref}
\hypersetup{
    colorlinks=true,
    linkcolor=blue,
    filecolor=black,      
    urlcolor=blue,
    citecolor=blue,
}
\usepackage{natbib}
\usepackage{braket}
\usepackage{soul}
\usepackage{relsize}
            
\usepackage{pifont}

\begin{document}
\draft
\title{State-dependent friction for a moving liquid contact line over rough solid surfaces}
\author{Caishan Yan}
\affiliation{Department of Physics, Hong Kong University of Science and Technology, Clear Water Bay, Kowloon, Hong Kong}
\author{Penger Tong}
\email{penger@ust.hk }
\affiliation{Department of Physics, Hong Kong University of Science and Technology, Clear Water Bay, Kowloon, Hong Kong}

\author{Qin Xu}
\email{qinxu@ust.hk }
\affiliation{Department of Physics, Hong Kong University of Science and Technology, Clear Water Bay, Kowloon, Hong Kong}

\date{\today}
\begin{abstract}
Solid friction between two rough surfaces is often observed to increase logarithmically over time due to contact creeping. An intriguing question is whether a similar aging effect occurs in contact line (CL) friction over rough substrates. Here, we report a systematic experimental study of CL friction using a hanging-fiber atomic force microscope (AFM) to measure the frictional force as a liquid CL moves across a fiber surface with different coatings under a well-controlled time protocol. State– (or time–)dependent CL friction is observed for the fiber surface with different textures in both the advancing and receding directions. The experimental findings are explained by a phenomenological model that links mesoscale CL friction to the microscopic relaxation of metastable air bubbles or liquid droplets trapped in the interstices of a rough surface. This model offers a general aging mechanism relevant to a wide range of liquid-solid interfaces. 
\end{abstract}
\maketitle

Liquid droplets traversing solid substrates are ubiquitous in both natural and industrial contexts, from raindrops cascading down window panes to various coating processes employed in manufacturing. A comprehensive understanding of droplet motion on diverse rough surfaces is crucial, as this motion is primarily governed by the pinning and depinning dynamics of the moving contact line (CL) between the liquid-air interface and the solid substrate, which may exhibit surface roughness and/or chemical heterogeneity \cite{Gennes85,Leger92,Bonn09,Snoeijer03}. This knowledge is essential for a wide array of engineering applications, including self-cleaning surfaces~\cite{Jiang2012}, microfluidic devices~\cite{Yu2023}, and innovative water collection systems~\cite{Wang2023}.

Our understanding of CL friction over rough surfaces primarily relies on the thermal activation model, which treats the depinning of a CL as a force-assisted barrier-crossing process~\cite{Blake06,Prevost99,Guan16}. This model effectively predicts the logarithmic (rate) dependence of the depinning force on the moving CL speed ($u$) \cite{Blake06,Prevost99,Guan16}. 
However, CL dynamics over rough surfaces reveal additional features that extend beyond the thermal activation framework. These include phenomena such as stick-slip instabilities~\cite{Doussal09,Yan24}, lubrication-assisted depinning~\cite{Lepikko23}, and asymmetric contact angle hysteresis~\cite{Guan16}. Such complexities highlight the intricate interplay of surface roughness and chemical heterogeneity. Recent studies have illuminated some intriguing similarities between CL friction and solid friction between two rough surfaces \cite{Li23}. 
For example, both types of friction exhibit logarithmic rate (or speed) dependence~\cite{Prevost99,Guan16,Li11} and demonstrate a transition from static to kinetic friction \cite{Gao17,Li11b,Yan23}. Solid friction has been extensively investigated \cite{Dieterich72, Dieterich78, Heslot94} and is known to exhibit state (or time) dependence due to contact creeping. However, it remains an intriguing question whether a similar aging effect manifests in CL friction. 

\begin{figure}
\centering
\includegraphics[width=0.4\textwidth]{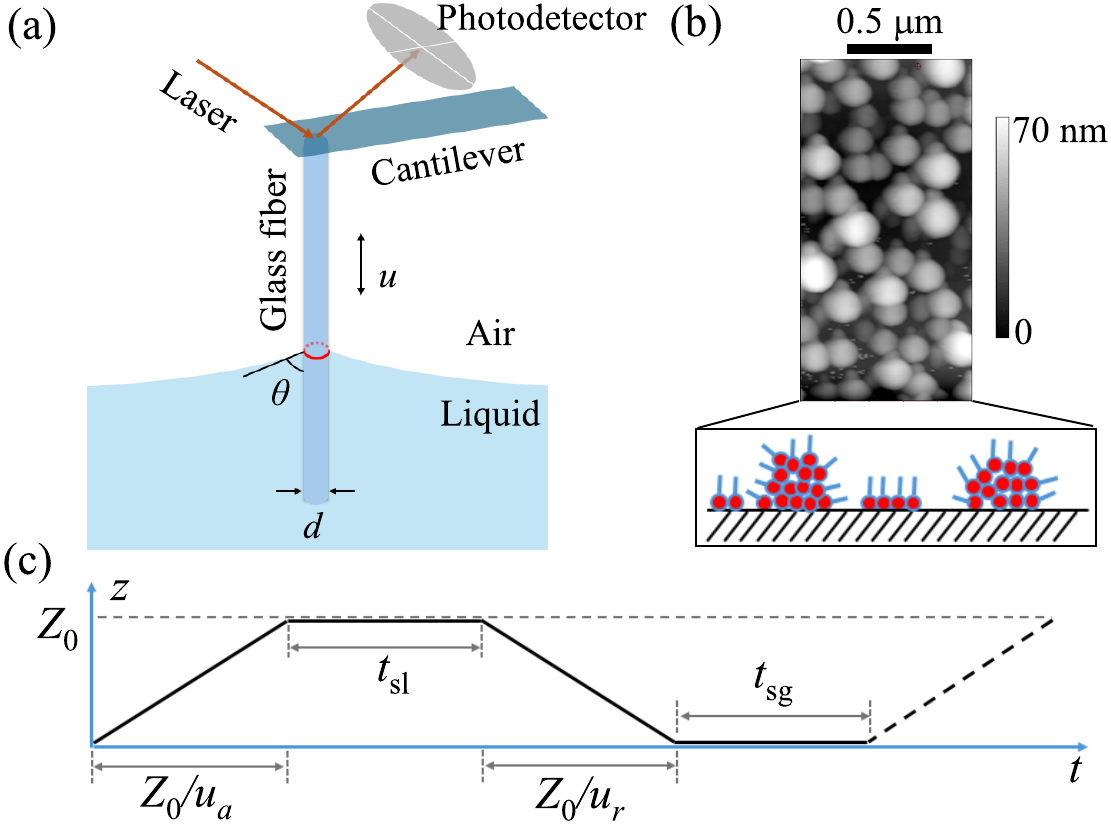}
\caption{{\bf Experimental setup.} { (a)} Schematic of the hanging-fiber AFM probe for measuring the capillary force acting on a circular CL (red line) formed at the intersection between the liquid-air interface and hanging glass fiber. (b) AFM topographical image of the PTS-coated fiber surface, with a magnified view illustrating the fiber surface coverage by PTS clusters, monolayer patches, and uncovered holes. (c) Fiber movement protocol, advancing-pause-receding-pause, with four control parameters, $u_a$, $t_{\rm sl}$, $u_r$ and $t_{\rm sg}$. }
\label{fig1}
\end{figure}

In this Letter, we present an experimental investigation of state-dependent CL friction on three different fiber surfaces with varying levels of physical roughness and chemical heterogeneity. As shown in Fig.~\ref{fig1}, a hanging-fiber AFM is used to measure the frictional force on a circular CL as it moves across the fiber surface under a controlled protocol during advancing and receding movements. The observed logarithmic increase in CL friction on different fiber surfaces is explained by a theoretical model that links the time-dependent CL friction to the dynamic wetting relaxation of metastable air bubbles or liquid droplets entrained by surface roughness. This model introduces a new aging mechanism that is totally different from the viscoelastic relaxation of CL on compliant substrates \cite{Guan20}, making it relevant to a wide range of liquid-solid interfaces with varying degrees of heterogeneity and roughness \cite{Parker01,Taylor2011,Wang16}.

{\em Experimental methods.} Figure~\ref{fig1}(a) shows the working principle of the hanging-fiber AFM. A thin glass fiber with a diameter $d\simeq 3~\mu$m and a length $\ell \simeq 250~\mu$m is glued to the free end of an AFM cantilever. The glass fiber is immersed in a 1~\% w/w propyltrichlorosilane (PTS) solution as the PTS molecules form covalent bonds to the glass substrate. After reaction, the resulting surface is covered by a layer of randomly distributed hydrophobic PTS aggregates with a typical size $\sim$200 nm, as shown in Fig.~\ref{fig1}(b) (see Ref.~\cite{Yan24} and Supplementary Information (SI) Sec.~I.A for more details \cite{Note1}). When the fiber is partially dipped into the liquid (de-ionized water), a circular CL forms at the intersection between the liquid-air interface and PTS-coated fiber surface. As the fiber moves vertically through the liquid bath, the capillarity force acting on the CL is measured by AFM. The fiber moving speed is kept at $u\leq 100$~$\mu$m/s, so that the corresponding capillary number is ${\rm Ca} \equiv \eta u/\gamma \leq 10^{-6}$, where $\eta$ is the fluid viscosity and $\gamma$ is the liquid-air surface tension. In this case, the AFM measures the capillarity force~\cite{Meglio90,Xiong,Guan16},
\begin{equation} 
f (z)= -\pi d \gamma \cos{\theta (z)},
\label{eq1}
\end{equation}
where $\pi d$ is the CL length, and $\theta(z)$ is the contact angle at vertical position $z$.  The sign of $f$ is defined as $f \leq 0$ for $\theta \leq 90^\circ$ and $f > 0$ for $\theta > 90^\circ$.

Figure~\ref{fig1}(c) illustrates the time protocol used to measure the capillarity force hysteresis (CFH) loop. The protocol consists of four consecutive steps: (i) The fiber moves downward at a constant speed $u_a$, producing an advancing capillary force $f_a(z)$ and corresponding advancing contact angle $\theta_a(z)$, as determined by Eq.~(\ref{eq1}). (ii) After reaching the maximum displacement $Z_0 = 28~\mu$m, the fiber is held stationary in water for a period $t_{sl}$. (iii) The fiber then moves upward at a receding speed $u_r$, producing a receding capillary force $f_r(z)$ and a receding contact angle $\theta_r(z)$. (iv) After reaching the starting position, the fiber is kept stationary with the PTS-coated surface being exposed to air for a period $t_{\rm sg}$ to complete the CFH loop. This CFH loop is repeated throughout the AFM measurements to ensure that the results are reproducible. We use the two pause times, $t_{\rm sg}$ and $t_{\rm sl}$, separately to control the initial state of the fiber surface prior to the advancing and receding movements (see SI Sec.~I.B for details \cite{Note1}).

\begin{figure}
\centering
\includegraphics[width=0.47\textwidth]{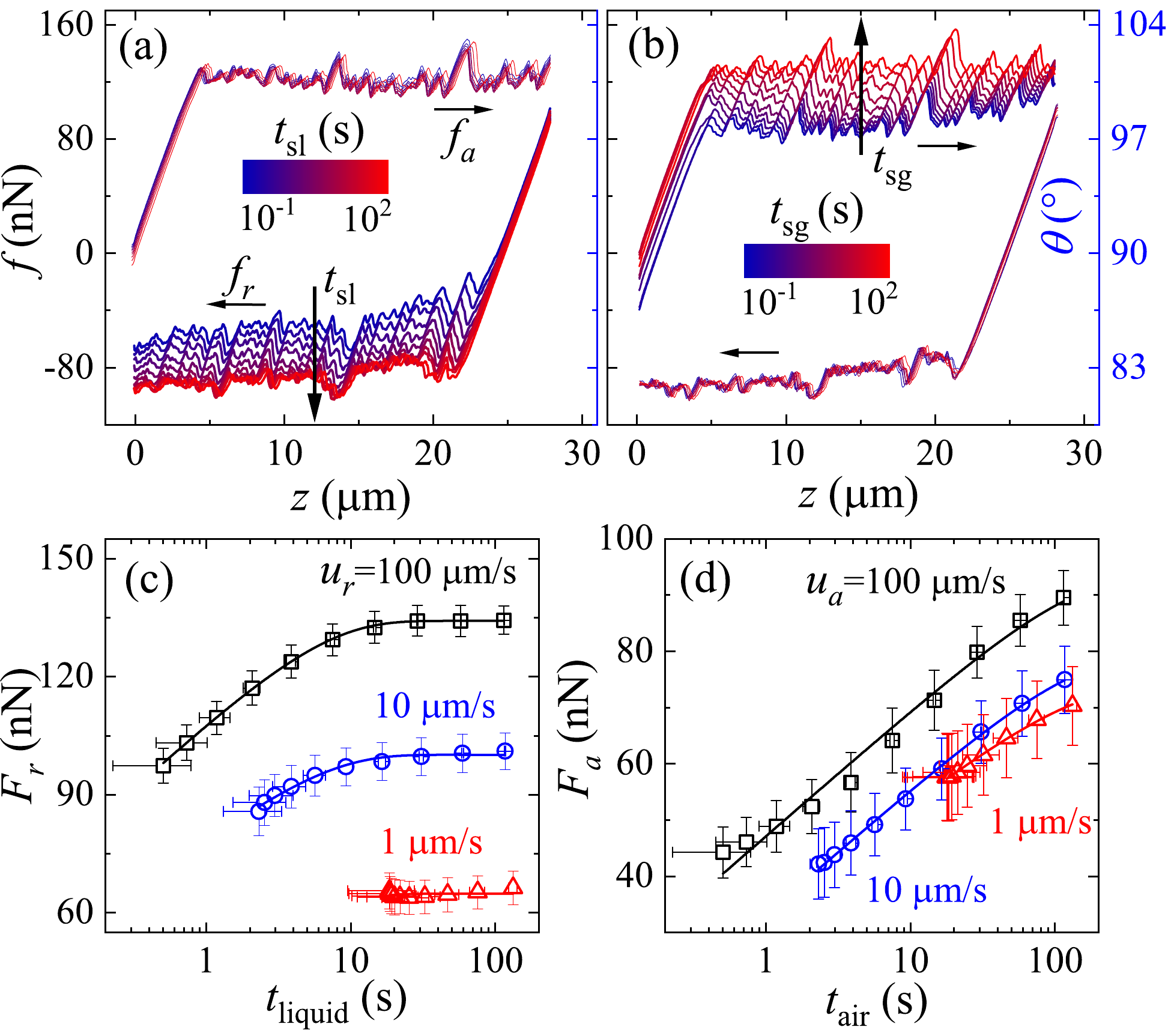}
\caption{{\bf Capillarity force hysteresis (CFH) on the PTS-coated fiber surface.} (a) (or (b)) Measured CFH loops for different values of $t_{\rm sl}$ (or $t_{\rm sg}$): 0.22, 0.45, 0.90, 1.79, 3.58, 7.17, 14.3, 28.7, 57.3, 114.7~s. The measurements are made when $t_{\rm sg}$ (or $t_{\rm sl}$) is held constant (= 20~s) and with $u_a = u_r =100$~$\mu$m/s. (c) Obtained receding friction $F_r$ as a function of $t_{\rm liquid}$ for $u_r=$~1, 10, and 100~$\mu$m/s. The solid lines show the best fits to Eq.~(\ref{eq6}) with $\Gamma_r = 5.7$~s. (d) Obtained advancing friction $F_a$ as a function of $t_{\rm air}$ for $u_a=$~1, 10, and 100~$\mu$m/s. The solid lines show the fits to Eq.~(\ref{eq6}) using $\Gamma_a = 230$~s.}
\label{fig2}
\end{figure}

{\em State-dependent CL friction.} Figures~\ref{fig2}(a) and \ref{fig2}(b) show the measured CFH loops on a PTS-coated fiber surface with $u_a=u_r=100$~$\mu$m/s, when $t_{\rm sg}$ and $t_{\rm sl}$ are varied separately. In each CFH loop, the measured capillary force in the advancing direction first increases linearly with $z$ because of the pinning of the CL on the fiber surface. A depinning transition occurs when the capillary restoring force of the stretched liquid-air interface exceeds the maximal pinning force, similar to the static-to-kinetic transition in solid–solid friction~\cite{Gao17}. After the depinning, the CL begins to move, giving rise to steady-state
fluctuations of $f_a(z)$ in the plateau region~\cite{Yan24}. A similar process is repeated in the receding direction. It is seen from Fig.~\ref{fig2}(a) that the measured $f_a(z)$ remains unchanged when $t_{\rm sg} $ is kept at constant, whereas the measured $f_r(z)$ in the receding direction gradually shifts downward with increasing $t_{\rm sl}$ from 0.22~s to 114.7~s. Similarly, Fig.~\ref{fig2}(b) reveals that the measured $f_r(z)$ remains unchanged when $t_{\rm sl}$ is kept at constant, whereas the measured $f_a(z)$ gradually shifts upward with increasing $t_{\rm sg} $ from 0.22~s to 114.7~s. These results indicate that the initial state of the water-PTS-coated-fiber interface changes continuously with $t_{\rm sg}$ and $t_{\rm sl}$ prior to the advancing and receding movements.

\begin{figure*}
\centering
\includegraphics[width=0.78\textwidth]{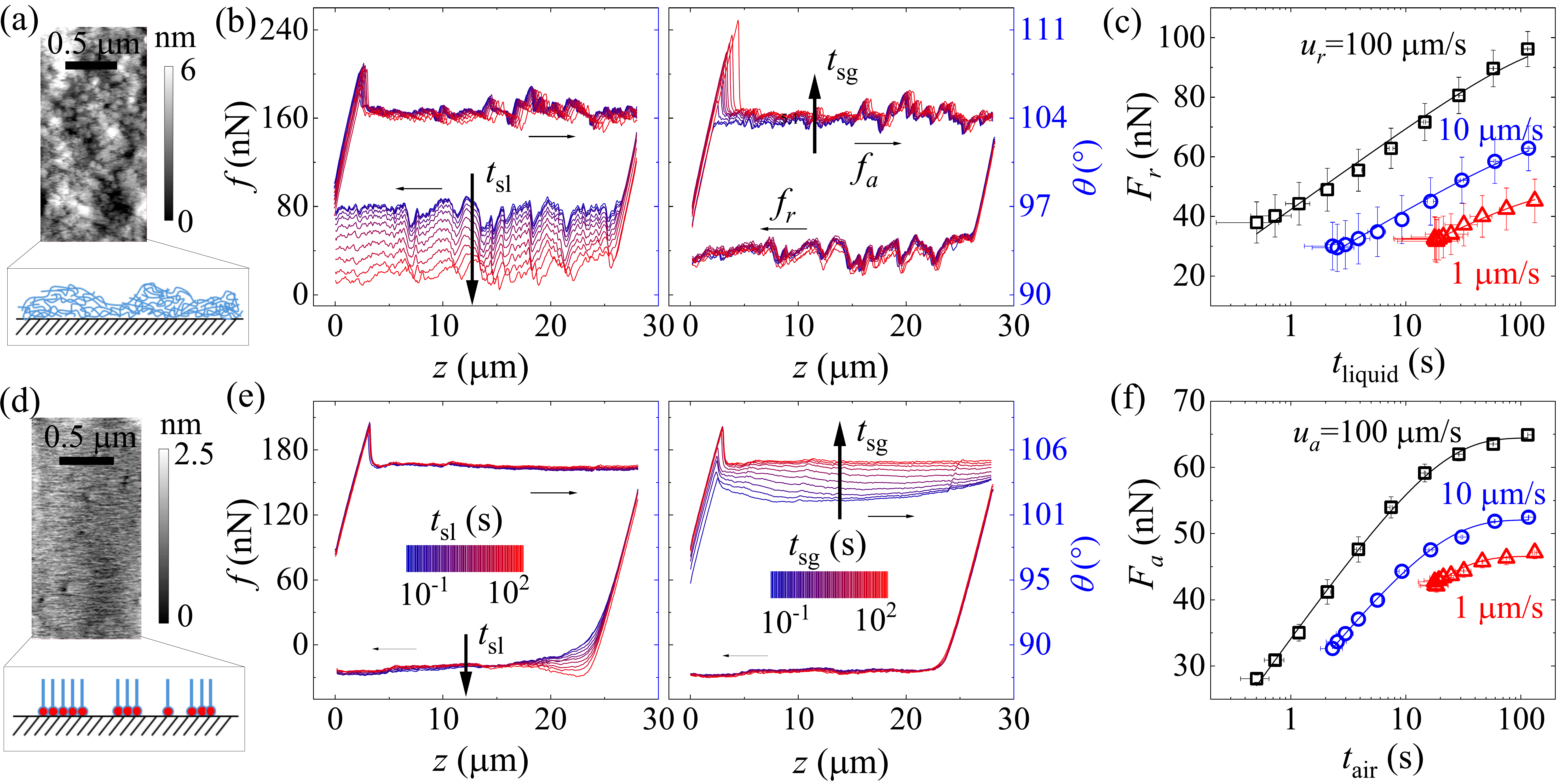}
\caption{{\bf Capillarity force hysteresis (CFH) on the PDMS- and FTS-coated fiber surfaces.} (a) \& (d) AFM topographical images of the (a) PDMS- and (d) FTS-coated fiber surfaces with a magnified view illustrating the surface profile of the (a) PDMS and (d) FTS layers. (b) \& (e) Measured CFH loops on the (b) PDMS- and (e) FTS-coated fiber surfaces with $u_a = u_r =100$~$\mu$m/s. The measurements are made for different values of $t_{\rm sl}$ (left panels) (or $t_{\rm sg}$ for the right panels): 0.22, 0.45, 0.90, 1.79, 3.58, 7.17, 14.3, 28.7, 57.3, 114.7~s, when $t_{\rm sg}$ (left panels) (or $t_{\rm sl}$ for the right panels) is held constant (= 20~s). (c) Obtained $F_r$ on the PDMS-coated fiber surface as a function of $t_{\rm liquid}$ for $u_r=$~1, 10, 100~$\mu$m/s. The solid lines show the best fits to Eq.~(\ref{eq6}) with $\Gamma_r = 230$~s. (f) Obtained $F_a$ on the FTS-coated fiber surface as a function of $t_{\rm air}$ for $u_r=$~1, 10, 100~$\mu$m/s.  The solid lines show the best fits to Eq.~(\ref{eq6}) with $\Gamma_a = 17.1$~s.}
\label{fig3}
\end{figure*}

When the fiber surface is ideally smooth and chemically homogeneous, one has $f_a=f_r=f_e$, where $f_a$ and $f_r$ are, respectively, the mean values of $f_a(z)$ and $f_r(z)$, and $f_e = - \pi \gamma d\cos{\theta_e} $ is the equilibrium value of the capillary force with $\theta_e$ being the equilibrium contact angle. The CL friction is actually the unbalanced capillary force due to surface roughness or chemical heterogeneity, i.e., 
$F_{i} = \vert f_i - f_e\vert$ with $i=(a, r)$,
where $f_i\equiv\langle f_i(z) \rangle$ is averaged over the plateau region (see SI Sec.~II.A for details about how $f_e$ is determined \cite{Note1}). Figure~\ref{fig2}(c) shows the obtained receding friction $F_r$ as a function of the CL's mean liquid contact time, $t_{\rm liquid}=(Z_0/2)/u_a+t_{\rm sl}+(Z_0/2)/u_r$ (see Fig.~\ref{fig1}(c)), for three different receding speeds $u_r$. Here, we use the maximal advancing speed $u_a=100~\mu$m/s accessible in the experiment to minimize the fiber approaching time, $(Z_0/2)/u_a= 0.14~s$. The contact time $t_{\rm liquid}$ is a measure of ``surface freshness" for the receding CL (see SI Sec.~I.B for more details \cite{Note1}). Similarly, Fig.~\ref{fig2}(d) exhibits the advancing friction $F_a$ as a function of the CL's mean air exposure time, $t_{\rm air}= (Z_0/2)/u_r+t_{\rm sg}+(Z_0/2)/u_a$ (see Fig.~\ref{fig1}(c)), for three different advancing speeds $u_a$. The maximal receding speed $u_r=100~\mu$m/s is employed to minimize the fiber withdraw time, $(Z_0/2)/u_r=0.14$~s.

As shown in Fig.~\ref{fig2}(c), for the two larger values of $u_r$ (100 and 10 $\mu$m/s), the obtained receding friction $F_r$ increases logarithmically over $t_{\rm liquid}$ and then saturates at a constant value when $t_{\rm liquid} \gtrsim 10$~s. For the slowest receding speed $u_r = 1$~$\mu$m/s, as the accessible liquid contact time $t_{\rm liquid}$ is always larger than 10~s, $F_r$ remains unchanged over the range of $t_{\rm liquid}$. In contrast, as shown in Fig.~\ref{fig2}(d), the resulting advancing friction $F_a$ for three different $u_a$ all increases logarithmically over $t_{\rm air}$ without reaching a saturation. These results suggest that the state-dependent CL friction on the PTS-coated surface occurs in both the advancing and receding directions. 

To test the universality of the observed state-dependent CL friction, we repeat the AFM measurements using glass fibers with two types of new coatings. One is dip-coated with a thin layer of polydimethylsiloxane (PDMS) gel with a 5 wt\% crosslinking density. Figure~\ref{fig3}(a) shows the AFM image of the PDMS-coated fiber surface, revealing a continuous hydrophobic coating with a Root Mean Square (RMS) roughness of 3 nm across a micron-sized area. The other is grafted with a monolayer of trichloro-(1H,2H,perfluorooctyl)-silane (FTS)~\cite{Guan16,Lepikko23}. The FTS-coated fiber surface is covered by a hydrophobic monolayer decorated with randomly distributed hydrophilic holes (coating defects) of glass surface (see Fig.~\ref{fig3}(d) and SI Sec.~I.A for details \cite{Note1}). 

As shown in Fig.~\ref{fig3}(b), for the PDMS-coated fiber surface, only the measured $f_r(z)$ reveals $t_{\rm sl}$-dependence, whereas the measured $f_a(z)$ remains invariant with $t_{\rm sg}$. As shown in Fig.~\ref{fig3}(c), the obtained receding friction $F_r$ with three values of $u_r$ all increases logarithmically over $t_{\rm liquid}$ without showing a saturation over the range of $t_{\rm liquid}$ studied. The FTS-coated fiber surface, on the other hand, exhibits an opposite trend. As shown in Fig.~\ref{fig3}(e), only the measured $f_a(z)$ reveals $t_{\rm sg}$-dependence, whereas the measured $f_r(z)$ remains invariant with $t_{\rm sl}$. The obtained advancing friction $F_a$ with three values of $u_a$ all increases logarithmically over $t_{\rm air}$ and then saturates at a constant value when $t_{\rm air} \gtrsim 30$~s, as shown in Fig.~\ref{fig3}(f). In addition, these robust features of the state-dependent CL friction over the three fiber surfaces are also observed when ethylene glycol is used as a wetting liquid (see SI Sec.~II.C for more details \cite{Note1}). 

{\em Phenomenological model.} The state-dependent CL friction over the three fiber surfaces is unlikely caused by the structural changes of the fiber surface under capillary pulling, because the surface coatings used are both physically and chemically stable. The surface stability is also certified by the high reproducibility of the repeated AFM measurements on the same fiber surface. Herein we consider a novel aging effect associated with the relaxation of a non-equilibrium coverage of entrained fluids (with metastable meso- or nano-scale air bubbles or liquid droplets) trapped on the rough fiber surface when it is quickly pushed downward or pulled upward through the liquid-air interface.  
\begin{figure}
\centering
\includegraphics[width=0.38\textwidth]{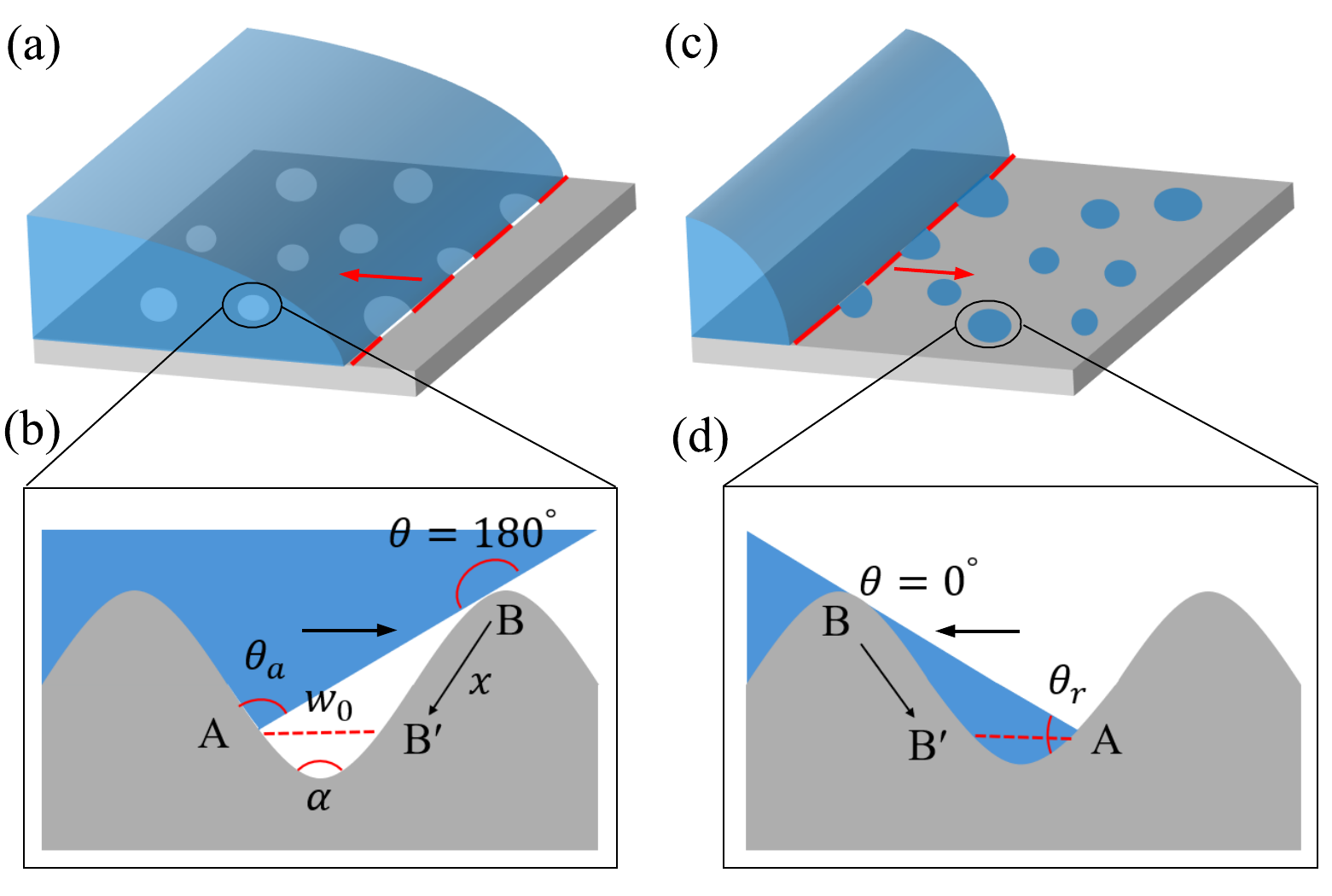}
\caption{{\bf Entrainment of air bubbles and liquid droplets by rough surfaces.}  (a) A receding CL moving over a hydrophobic fiber surface covered a layer of metastable air bubbles. 
(b) Formation and relaxation of a trapped air bubble by surface roughness.
(c) An advancing CL moving over a rough fiber surface covered by a layer of entrained liquid droplets. (d) Formation and relaxation of an entrained liquid droplet by surface roughness.}
\label{fig4}
\end{figure}
    
When a rough hydrophobic fiber is quickly pushed through the liquid-air interface, air bubbles tend to be trapped in the interstices of the rough elements. As a result, the liquid-contacted portion of the fiber surface is covered with a layer of isolated metastable air bubbles, as indicated in Fig.~\ref{fig4}(a). While the phenomenon of air-trapping on hydrophobic surfaces has been reported in previous studies \cite{Ishida00, Poetes10, Karpitschka2012, Christopher22}, the non-equilibrium nature of the air bubbles trapped by surface roughness has remained unknown.  As illustrated in Fig.~\ref{fig4}(b), when the advancing CL approaches Point $A$ near the bottom of a valley, the liquid-air interface can make a new contact with the solid substrate at Point $B$, if the advancing contact angle $\theta_a$ exceeds the opening angle $\alpha$ of the valley. Hence, an air bubble is trapped below Line $AB$ when $\theta_a > \alpha$. Because the initial contact angle at Point $B$ (close to 180$^\circ$) is larger than $\theta_a$, the bubble's interface will subsequently relax from $B$ to an equilibrium point $B^\prime$ under the unbalanced capillary force \cite{Wang16} 
\begin{equation}
f_{c}=\gamma\vert\cos\theta(x)-\cos\theta_{e}\vert\simeq (\gamma x/w_0) \cos{(\alpha/2)} \sin\theta_{e},
\label{eq2}
\end{equation}
where $w_0$ is a typical width of the bubble and $x$ represents the bubble's CL movement towards $B^\prime$ (see Fig.~\ref{fig4}(b)). 

As the air bubbles relax with the liquid contact time $t_{\rm liquid}$, their coverage on the fiber surface shrinks. Therefore, when the fiber is pulled back, the portion of the receding CL in contact with the air bubbles (see Fig.~\ref{fig4}(a)) will depend on $t_{\rm liquid}$. For small values of $t_{\rm liquid}$, the air bubbles have not relaxed much and a larger portion of the receding CL is in contact with the air bubbles, which do not contribute to the receding CL friction $F_r(t_{\rm liquid})$. For large values of $t_{\rm liquid}$, a smaller portion of the CL is in contact with the air bubbles. As a result, $F_r(t_{\rm liquid})$ will increase with $t_{\rm liquid}$, giving rise to a state-dependent receding CL friction.

Similarly, when a rough fiber surface with hydrophilic defects is quickly withdrew from the liquid-air interface, liquid droplets tend to be entrained by these hydrophilic domains, as indicated in Fig.~\ref{fig4}(c). As illustrated in Fig.~\ref{fig4}(d), a liquid droplet is entrained when the receding CL becomes less steep than the rough surface slope, i.e., $\theta_r < (\pi-\alpha)/2$. As the entrained liquid droplets relax with the air exposure time $t_{\rm air}$, the advancing CL encounters less liquid interface (see Fig.~\ref{fig4}(c)), which do not contribute to the advancing CL friction $F_a(t_{\rm air})$. Consequently, $F_a(t_{\rm air})$ will increase with $t_{\rm air}$, giving rise to a state-dependent advancing CL friction.

The relaxation of the trapped air bubbles or liquid droplets is slow, because their local CLs with the solid substrate encounter many pinning defects.  Assuming these local defects have a
typical size $\lambda$, energy barrier $E_b$ (the corresponding depinning force $\sim E_b/\lambda$), and line density $a/\lambda$ with $a\leq 1$ being a numerical constant, the depinning force needed to keep a droplet CL moving at a speed $\dot{x}$ is given by \cite{Bell78,Friddle2008,Guan16}
\begin{equation}
f_{d}\simeq 2(a/\lambda)f_T \ln(1-\dot{x}/u_0), 
\label{eq3}
\end{equation}
where $f_T=k_BT/\lambda$ is the thermal force, and $u_0$ is a thermal speed associated with the barrier-crossing over $E_b$~\cite{Guan16}. By equating $f_d$ with $f_c$ in Eq.~(\ref{eq2}), we find the local droplet CL movement, $x(t)=-\xi\ln(1-(1-e^{-x(0)/\xi})e^{-t/\tau})\simeq -\xi \ln (t/\tau)$, where $\xi$ is a characteristic length proportional to $w_0$ and $\tau = \xi/u_0$. The second equality is obtained when $x(0) \gg \xi$ and $t \ll \tau$. 
This result suggests that the local droplet CLs relax logarithmically over time $t$ from $x(0)$ to $0$. The total frictional force for an advancing (or receding) CL intersecting with $n$ independent fluid droplets can be written as (see SI Sec.~III for more details \cite{Note1})
\begin{equation}
F_i(t)\simeq (F_0)_i +\beta_i\ln(1-e^{-t/\Gamma_i}), \;\;\; (t>0;\; i=a, r),
\label{eq6}
\end{equation}
where $\beta_i$ and $\Gamma_i$ are two fitting parameters for a given fiber surface, and $(F_0)_i$ is the asymptotic value of $F_i(t)$ when the rough surface is fully relaxed ($t \gg \Gamma_i$). For $0<t\ll \Gamma_i$, we find the CL friction increases logarithmically with time $t$, $F_i(t) \simeq (F_0)_i + \beta_i \ln (t/\Gamma_i)$, in agreement with the experimental findings shown in Figs.~\ref{fig2} and~\ref{fig3}. 

{\em Discussions.} 
The solid lines in Figs.~\ref{fig2}(c), \ref{fig2}(d), \ref{fig3}(c), and \ref{fig3}(f) show the best fittings of Eq.~(\ref{eq6}) to the data points. The three coated fiber surfaces display distinct behaviors in state-dependent CL friction. The FTS-coated fiber surface is covered by a relatively smooth hydrophobic monolayer decorated with hydrophilic voids, which tend to entrain liquid droplets from a receding CL and thus only give rise to a state-dependent advancing CL friction. In contrast, the PDMS-coated fiber surface is fully covered by a thin layer of relatively rough and stiff hydrophobic gel, which tends to trap air bubbles from a advancing CL and thus only produces a state-dependent receding CL friction. The PTS-coated fiber surface is covered by both hydrophobic clusters and hydrophilic voids and thus is capable of producing state-dependent advancing and receding CL frictions (see SI Sec.~I.A for more details \cite{Note1}). The actual relaxation rate for a CL intersecting with many trapped fluid droplets depends on the details of the surface texture, and the measured values of $\Gamma_a$ and $\Gamma_r$ provide a sensitive signature of the local surface roughness and heterogeneity. Our work thus reveals a novel mechanism for state-dependent CL friction, involving pinning and depinning dynamics over a landscape of random defects and roughness.

{\em Acknowledgement.} This work was supported by the Hong Kong RGC under grant nos.~16305821(Q.X.), 16306723(Q.X.), 16300421 (P.T.), 16300224 (P.T.), and C6041-24G (Q.X. \& P.T.). C.Y. acknowledges the support by the Hong Kong ITF-RTH Fellowship.


\begin{thebibliography}{99}
\bibitem{Gennes85}{P.-G. de Gennes, Wetting: statics and dynamics, Rev. Mod. Phys. {\bf 57}, 827 (1985).}
\bibitem{Leger92}{L. Leger and J.-F. Joanny, Liquid spreading, Rep. Prog. Phys. {\bf 55}, 431 (1992).}
\bibitem{Bonn09}{D. Bonn, J. Eggers, J. Indekeu, J. Meunier and E. Rolley, Wetting and spreading, Rev. Mod. Phys. {\bf 81}, 739, (2009).}
\bibitem{Snoeijer03}{J. Snoeijer and B. Andreotti, Moving contact lines: Scales, regimes, and dynamical transitions, Annu. Rev. Fluid Mech. {\bf 45}, 269 (2013).}
\bibitem{Jiang2012}{K. Liu, and Lei, Jiang, Bio-inspired Self-Cleaning Surfaces, \href{https://www.annualreviews.org/content/journals/10.1146/annurev-matsci-070511-155046}{Annu. Rev. Mater. Res. {\bf 42}, 231-263 (2012).}}
\bibitem{Yu2023}{Y. Qiu, K. Xu, A. A. Pahlavan, and R. Juanes, Wetting transition and fluid trapping in a microfluidic fracture, \href{https://www.pnas.org/doi/abs/10.1073/pnas.2303515120}{Proc. Natl. Acad. Sci. {\bf 120(22)}, e2303515120 (2023).}}
\bibitem{Wang2023}{Y. Wang, W. Zhao, M. Han, J. Xu, and K. C. Tam, Biomimetic surface engineering for sustainable water harvesting systems, \href{https://www.nature.com/articles/s44221-023-00109-1}{Nat. Water. {\bf 1}, 587-901 (2023).}}
\bibitem{Blake06}{Blake, T. D., The physics of moving wetting lines, \href{https://www.sciencedirect.com/science/article/pii/S0021979706002463}{J. Colloid Interface Sci. {\bf 299}, 1-13 (2006).}}
\bibitem{Prevost99}{Prevost, A., Rolley, E., Guthmann, C., Thermally activated motion of the contact line of a liquid $^4$He meniscus on a cesium substrate, \href{https://journals.aps.org/prl/abstract/10.1103/PhysRevLett.83.348}{Phys. Rev. Lett. {\bf 83}, 348 (1999).}} 
\bibitem{Guan16}{Guan, D., Wang, Y. J., Charlaix, E., Tong, P., Asymmetric and speed-dependent capillary force hysteresis and relaxation of a suddenly stopped moving contact line, \href{https://journals.aps.org/prl/abstract/10.1103/PhysRevLett.116.066102}{Phys. Rev. Lett. {\bf 116}, 066102 (2016).}} 
\bibitem{Doussal09}{Doussal., P., Wiese. K. J., Moulinet., S., and Rolley. E., Height fluctuations of a contact line: A direct measurement of the renormalized disorder correlator, \href{https://iopscience.iop.org/article/10.1209/0295-5075/87/56001/meta}{EPL {\bf 87}, 56001 (2009).}}
\bibitem{Yan24}{Yan, C., Guan, D., Wang, Y., Lai, P. Y., Chen, H. Y., Tong, P., Avalanches and extreme value statistics of a mesoscale moving contact line, \href{https://journals.aps.org/prl/abstract/10.1103/PhysRevLett.132.084003}{Phys. Rev. Lett. {\bf 132}, 084003 (2024).}} 
\bibitem{Lepikko23}{S. Lepikko, Y. M. Jaques, M. Junaid, M. Backholm, J. Lahtinen, Jaakko Julin, Ville Jokinen, T. Sajavaara, M. Sammalkorpi, A. S. Foster, R. H. A. Ras, Droplet slipperiness despite surface heterogeneity at molecular scale, \href{https://www.nature.com/articles/s41557-023-01346-3}{Nat. Chem., {\bf 16}, 506-513 (2024).}}
\bibitem{Li23}{Li, X., Bodziony, F., Yin, M., Marschall. H., Berger. R., and Butt. H.-J., Kinetic drop friction, \href{https://www.nature.com/articles/s41467-023-40289-8}{Nat. Commun. {\bf 14}, 4571 (2023).}}
\bibitem{Li11}{Li, Q., Dong, Y., Perez, D., Martini, A., and Carpick, R. W., Speed dependence of atomic stick-slip friction in optimally matched experiments and molecular dynamics simulations, \href{https://journals.aps.org/prl/abstract/10.1103/PhysRevLett.106.126101}{Phys. Rev. Lett. {\bf 106}, 126101 (2011).}}
\bibitem{Li11b}{Li, Q., Tullis, T. E., Goldsby, D., and Carpick, R. W., Frictional ageing from interfacial bonding and the origins of rate and state friction, \href{https://www.nature.com/articles/nature10589}{Nature {\bf 106}, 126101 (2011).}}
\bibitem{Gao17}{Gao, N., Geyer, F., Pilat, D. W., Wooh, S., Vollmer, D., Butt, H., Berger, R., How drops start sliding over solid surfaces, \href{https://www.nature.com/articles/nphys4305}{Nat. Phys. {\bf 14}, 191-196 (2018).}}
\bibitem{Yan23}{Yan, C., Chen, H. Y., Lai, P. Y., and Tong, P., Statistical laws of stick-slip friction at mesoscale, \href{https://www.nature.com/articles/s41467-023-41850-1}{Nat. Commun. {\bf 14}, 6221 (2023).}} 
\bibitem{Dieterich72}{Dieterich, J. H., Time-dependent friction in rocks, \href{https://agupubs.onlinelibrary.wiley.com/doi/abs/10.1029/JB077i020p03690}{J. Geophys. Res. {\bf 77}, 20 (1972).}}
\bibitem{Dieterich78}{Dieterich, J. H., Time-dependent friction and the mechanics of stick-slip, \href{https://link.springer.com/article/10.1007/BF00876539}{Pure Appl. Geophys., {\bf 116}, 790-806 (1978).}} 
\bibitem{Heslot94}{Heslot, F., Baumberger, T., Perrin, B., Caroli, B., and Caroli, C., Creep, stick-slip, and dry-friction dynamics: Experiments and a heuristic model, \href{https://journals.aps.org/pre/abstract/10.1103/PhysRevE.49.4973}{Phys. Rev. E {\bf 49}, 4973 (1994).}}
\bibitem{Guan20}{Guan, D., Charlaix, E., Tong, P., State and rate dependent contact line dynamics over an aging soft surface, \href{https://journals.aps.org/prl/abstract/10.1103/PhysRevLett.124.188003}{Phys. Rev. Lett. {\bf 124}, 188003 (2020).}} 
\bibitem{Parker01}{Parker, A. R. and Lawrence, C. R., Water capture by a desert beetle, \href{https://www.nature.com/articles/35102108}{Nature {\bf 414}, 33–34 (2001).}}
\bibitem{Taylor2011}{P. Taylor, The wetting of leaf surfaces, \href{https://www.sciencedirect.com/science/article/pii/S135902941000124X}{Curr. Opin. Colloid Interface Sci., {\bf 4}, 326-334 (2011).}}
\bibitem{Wang16}{Wang, Y. J., Guo, S., Chen, H. Y., Tong, P., Understanding contact angle hysteresis on an ambient solid surface, \href{https://journals.aps.org/pre/pdf/10.1103/PhysRevE.93.052802}{Phys. Rev. E {\bf 93}, 052802 (2016).}}  
\bibitem{Note1}
{See Supplemental Information at http://... which complement the main text.}
\bibitem{Meglio90}{Di Meglio, J.-M., Qu{\'e}r{\'e}, D., Contact angle hysteresis: a first analysis of the noise of the creeping motion of the contact line, \href{https://iopscience.iop.org/article/10.1209/0295-5075/11/2/012/meta}{Europhys. Lett. {\bf 11}, 163 (1990).}} 
\bibitem{Xiong}{Xiong, X.-M., Guo, S., Xu, Z.-L., Sheng, P., Tong, P., Development of an atomic-force-microscope-based hanging-fiber rheometer for interfacial microrheology, \href{https://journals.aps.org/pre/abstract/10.1103/PhysRevE.80.061604}{Phys. Rev. E {\bf 80}, 061604 (2009).}} 
\bibitem{Ishida00}{Ishida, N., Inoue, T., Miyahara, M., Higashitani, K., Nano bubbles on a hydrophobic surface in water observed by tapping-mode atomic force microscopy, \href{https://pubs.acs.org/doi/10.1021/la000219r}{Langmuir {\bf 16}, 6377-6380 (2000).}} 
\bibitem{Poetes10}{Poetes, R., Holtzmann, K., Franze, K., Steiner, U., Metastable underwater superhydrophobicity, \href{https://journals.aps.org/prl/abstract/10.1103/PhysRevLett.105.166104}{Phys. Rev. Lett. {\bf 105}, 166104 (2010).}} 
\bibitem{Karpitschka2012}{S. Karpitschka, E. Dietrich, J. R.T.Seddon, H. J.W.Zandvliet, D. Lohse, and H. Riegler, Nonintrusive Optical Visualization of Surface Nanobubbles, \href{https://journals.aps.org/prl/abstract/10.1103/PhysRevLett.109.066102}{Phys. Rev. Lett. {\bf 109}, 066102 (2012).}}
\bibitem{Christopher22}{Vega-Sanchez, C., Peppou-Chapman, S., Zhu, L., Neto, C., Nanobubbles explain the large slip observed on lubricant-infused surfaces, \href{https://www.nature.com/articles/s41467-022-28016-1}{Nat. Commun. {\bf 13}, 1-11 (2022).}} 
\bibitem{Bell78}{Bell, G. I., Models for the specific adhesion of cells to cells. Science {\bf 200}, 618 (1978).}
\bibitem{Friddle2008}{Friddle,R. W., Unified model of dynamic forced barrier crossing in single molecules. Phys. Rev. Lett. {\bf 100}, 138302 (2008).}
        
\end{thebibliography}
\end{document}